\documentstyle[eqsecnum,amssymb,prb,aps,multicol,psfig]{revtex}

\begin{document}
\draft
\title{Crater formation by fast ions: \\
comparison of experiment with Molecular Dynamics simulations}
\author{E. M. Bringa and R. E. Johnson}
\address{Engineering Physics, University of Virginia, Charlottesville VA 22903 U.S.A.}
\author{R. M. Papal\'{e}o}
\address{Faculdade de F\'{\i }sica, Universidade Cat\'{o}lica do Rio Grande do Sul, \\
CP 1429, Porto Alegre, RS, Brazil}
\date{\today}
\maketitle

\begin{abstract}
An incident fast ion in the electronic stopping regime produces a track of
excitations which can lead to particle ejection and cratering. Molecular
Dynamics simulations of the evolution of the deposited energy were used to
study the resulting crater morphology as a function of the excitation
density in a cylindrical track for large angle of incidence with respect to
the surface normal. Surprisingly, the overall behavior is shown to be
similar to that seen in the experimental data for crater formation in
polymers. However, the simulations give greater insight into the cratering
process. The threshold for crater formation occurs when the excitation
density approaches the cohesive energy density, and a crater rim is formed
at about six times that energy density. The crater length scales roughly as
the square root of the electronic stopping power, and the crater width and
depth seem to saturate for the largest energy densities considered here. The
number of ejected particles, the sputtering yield, is shown to be much
smaller than simple estimates based on crater size unless the full crater
morphology is considered. Therefore, crater size can not easily be used to
estimate the sputtering yield.
\end{abstract}


%% Draft adds the pack numbers

\begin{multicols}{2}
%\twocolumn
\narrowtext

\section{Introduction}

Surface modification of materials by single ion irradiation has been studied
in insulators \cite{kopni,eriksson,eriksson1,ricardoprb,papaleo,balanzat}, 
semiconductors \cite{averbackSi,nordlund11,bumps} and metals \cite{dammak,bircher} 
using electron, STM and AFM microscopes. 
A large variety of features have been observed: `bumps' \cite{bumps}, craters 
\cite{kopni,eriksson,eriksson1,ricardoprb,papaleo,reimann-crater,vorobyova,urba-crater}, 
crater rims (hillocks) \cite{kopni,eriksson,eriksson1,ricardoprb,papaleo,muller,neumann}, 
adatoms \cite{nordlund11,nordlund}, and surface roughening \cite{insepov-rough}. 
Cratering occurs in response to the pressure pulse and fluid
flow to the surface produced by the rapid deposition of energy, but the
process is not understood quantitatively. `Bumps' generally appear when an
energetic process occurs a few layers below the surface creating a low
density region with a larger volume which raises the surface. \cite
{averback1}. When the energy loss per unit path length of the
projectile, $dE/dx$, and\ the sputtering yield are relatively small, adatoms
are observed in both experiments and simulations. By increasing the energy
deposition (and the yield) craters are eventually formed. For very large
energy deposition and yields, re-deposition of the ejecta plus plastic
deformation occurs, producing craters with rims, studied recently for ion
bombardment of polymers and other organic materials \cite{papaleo}.

Craters are also produced by cluster ion bombardment which can lead to huge
sputtering yields \cite{andersen,andersen-angle}. This process has
been studied in the velocity regime in which nuclear (elastic) energy loss
dominates over electronic energy loss and has been seen in both experiments 
\cite{andersen,yamada2} and simulations %
\cite{shulga-clusters,averback-clusters1,insepov,insepov1}. 
The simulations are generally performed for bombardment at normal
incidence and when energy is deposited in momentum transfer collisions to
the target atoms. There are few simulations of cratering in the electronic
regime \cite{david} and none for non-normal incidence.

At normal incidence the crater produced by a fast incident ion has a roughly
circular profile, but recent experiments have focused on ions incident at a
large angle with respect to the normal \cite{papaleo} and at grazing 
incidence \cite{vorobyova}. Even at normal incidence MD results for keV ion 
clusters incident on a copper surface appear to disagree with the scaling 
laws followed by macroscopic cratering \cite{urba-crater} in which the crater
radius varies with the bombarding energy \cite{moon} $E$ as $E^{1/3}$. 
For oblique incidence in polymers \cite{papaleo} and biomolecules \cite
{eriksson1} bombarded by fast heavy ions it was found that the crater width
does not increase significantly with increasing deposited energy density
whereas the size of the crater along the incident ion direction increases
rapidly with increasing energy density.

In this paper molecular dynamics (MD) simulations are used to study the
surface morphology produced by the energy deposited by fast ions incident at
large angle with respect to the surface normal. The results of these
simulations are compared to models for the length, width and depth of the
crater vs. the energy density (i.e., $dE/dx$ and track width). Since crater
formation is used for sculpting specific surfaces features for biomolecule
adsoption \cite{quist}, for determining surface properties \cite{ricardoprb}%
, and for estimating sputtering yields \cite{papaleo,eriksson1}, we use MD
simulations to extract scaling laws for crater formation. Although the
simulations are for an ``atomic'' solid, quite remarkably, the trends are
very similar to those recently seen in polymers \cite{papaleo}. However, in
the simulations we can study how cratering depends on the material 
properties and on the energy density deposited by an incident ion.

\section{MD simulation}

Following the passage of a fast heavy ion a cylindrically energized region
is produced in a solid, which we refer to as a track of excitations. A
Lennard-Jones (L-J) crystalline solid is simulated with particles
interacting through the potential \cite{paperI} $V\left( r\right)
=4\varepsilon \left[ \left( r/\sigma \right) ^{6}-\left( r/\sigma \right)
^{12}\right] $. Although this is an oversimplified model of a real solid,
this two parameter potential has the advantage that the equations of motion,
and, hence, all results, including the crater dimensions, scale with $%
\varepsilon $ and $\sigma $. In addition, certain weakly bound solids, such
as the low-temperature, condensed-gas solids, can be reasonably approximated
as L-J solids, with parameters $\varepsilon $ and $\sigma $ taken to
reproduce the material properties. All L-J samples have a cohesive energy $%
U\approx 8\varepsilon $. The interlayer distance for (001) layers is $%
d\approx 0.78\sigma $ and the bulk modulus is $B=75\varepsilon /\sigma ^{3}$%
. More details on the MD simulation can be found elsewhere \cite
{paperI,paperII,angle}. As in our earlier papers, the scaling with $%
\varepsilon $ and $\sigma $ is replaced by scaling using $U$ and the number
density $n$. For the fast processes which determine sputtering and cratering 
we showed that the scaling was roughly preserved when a more complex potential 
was used \cite{prb}.

Since the results will be compared to data on polymers we note that certain
polymers are roughly simulated using a L-J potential for the inter-chain
interactions, plus a stronger potential to account for the covalent
interaction within the chain \cite{binder}. A typical size for $\sigma $ in
polymers is 3.5-5 \AA\ \cite{binder}. The cohesive energy of a polymer is
more difficult to define. The covalent bonds among atoms in the same polymer
chain are of the order of several eV's with slightly weaker bonds between
monomers in a chain. However, the bonding among atoms in the neighboring
chains is very weak, much smaller than 1 eV, making the binding field
``anisotropic''. Removing a small chain requires different energies
depending on the chain orientation and entanglement. The average cohesive
energy is usually taken to be equal to the sublimation energy. A simple
estimate \cite{binding} gives $U\approx 0.5$ eV/monomer. Therefore, even
though the L-J calculations scale with size and binding energy, we assume an
effective binding energy of 0.5 eV/particle and $\sigma =$5 \AA\ so that our
``atoms'' very crudely represent monomers. The mass of the simulated
particle only changes the time scale, which is given by the dimensionless
time $t/t_{o}$, where $t_{o}=\sigma \sqrt{M/\varepsilon }$, and $M$ is the
mass of the simulated particle. Assuming a mass of $70u$ gives $t_{o}=1.75$
ps.

The stopping power, i.e. the energy deposited per unit length $dE/dx$, and
the track radius, here $r_{cyl}$, are typically used to describe the energy
density deposited by the ion in its passage through the solid. Since in the
electronic sputtering regime only a fraction of the experimental $dE/dx$
goes into non-radiative de-excitations, in the following we use the symbol $%
\left( dE/dx\right) _{eff}$ to represent the amount of energy deposition
contributing to track formation, cratering and sputtering. This fraction has
been estimated to be $\sim 0.2$ for MeV He$^{+}$ bombardment of solid O$_{2}$
\cite{o2}. We also use this here to compare the experimental results with
our MD simulations. To mimic the non-radiative energy release at the ion
track in the MD simulations all $N_{exc}$ particles within a cylinder of
radius $r_{cyl}$ are given an energy $E_{exc}$ with their velocities in
random directions. Therefore, $\left( dE/dx\right) _{eff}$ is $%
N_{exc}E_{exc}/d$. A track radius can be estimated from the Bohr adiabatic
radius \cite{kinetics}, though this has been questioned recently \cite
{o2,signew}. Thermal spike models have been applied recently to estimate the
latent track radius in irradiated polymers with positive results \cite
{szenes-poly}. In this paper all simulations were run for an initial track
radius $r_{cyl}=2\sigma =10$ \AA , which implies $N_{exc}\approx 10/\cos
\Theta $. The incident angle, $\Theta $, is measured in degrees with respect
to the surface normal. For all simulations $\Theta =60$. Larger angles were
not feasible as the sample size required became too large to practically 
simulate. Therefore, the stopping power from MD is multiplied by a factor 
$\cos 60/\cos 79$ to compare to experiments
done at $\Theta =79$. The size of the simulated sample was varied depending
on the size of $E_{exc}$ such that the results did not depend on the
boundary conditions used, and the total simulation time was also varied to
be able to ``detect'' all ejected particles. Most quantities presented are
averages of results from a number of simulations in which the directions of
the energized atoms were randomly varied.

\section{Crater features}

\begin{figure}[tbp]
\centerline{\psfig{figure=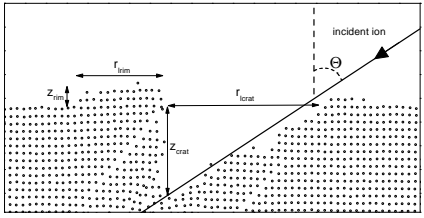,width=6.5cm,height=5cm,angle=0}}
\caption{ Slice of the final configuration of one MD cratering event,
showing a side view of the crater and crater rim. The incident ion impact
angle is $\Theta =60$, and $\left( dE/dx\right)_{eff}\approx 330U/\sigma =
330$ eV/nm. The different dimensions of the defect are shown in the figure.}
\label{fig1}
\end{figure}

A cut-across a crater formed following an excitation event is shown in Fig. 
\ref{fig1}. This cut is in a plane containing the initial surface normal and
the track direction and shows the maximum depth of the crater. The crater
wall is seen to have a slope similar to the incident ion direction on the
entrance side and a very steep slope at the back. Remarkably, this shape is
very stable even in this model solid as we have increased the run time by a
factor of two and see little relaxation occurring. This is the case because
of the rapid cooling in the track core discussed elsewhere \cite
{paperII,prb-new}. At the end of a typical simulation ($\sim 30t_{o}$) the
temperature is $\sim $50 K higher than the initial temperature of the sample
(30 K). Of course, over very long times, relaxation can occur in an atomic
material even at relatively low temperatures, whereas it is less likely in a
polymer. Therefore, crater morphology in polymers can be readily studied.
Despite some problems with their interpretation, scanning force microscopy
measurements \cite{neumann} are regularly used to analyze heavy ion damage
in solids.

\begin{figure}[tbp]
\centerline{\psfig{figure=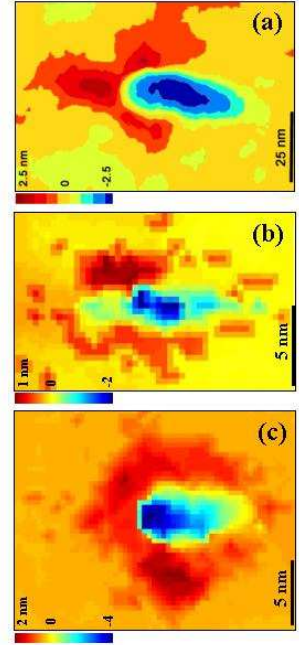,width=7.9cm,height=10.53cm,angle=0}}%width=13cm,height=17.3cm
\caption{Top view of a crater, where the color scale indicates height.
(a) experimental result for $\Theta =79$, and %
$\left(dE/dx\right)_{eff}=660$ eV/nm (20 MeV Au on PMMA). 
(b) MD simulation for $\Theta =75$ and $\left( dE/dx\right) _{eff}=205$
eV/nm. (c) MD simulation for $\Theta =60$ and $\left( dE/dx\right)_{eff}=330$
eV/nm. Because of differences in $\Theta$ and energy, the MD crater in (c) is 
not as elongated as the experimental crater.}
\label{fig2}
\end{figure}

Since all craters were found to have a characteristic morphology, the
dimensions used in the subsequent graphs and discussions are indicated in
Fig. \ref{fig1}. This crater is formed at high excitation densities ($12nU$)
and exhibits a rim on the back side. Not all craters, however, have rims, as
we will discuss. To `see' the rim better, in Fig. \ref{fig2} we show a top 
view of a crater \cite{web}, together with an AFM image of a heavy ion 
impacting on a PMMA film. The rim is primarily on the sides and the back of the 
crater, not at the ion entrance site. This is the case even when there is no 
momentum preferentially deposited along the incident angle as is the case in
macroscopic cratering. Here the crater is formed from a cylindrical `heat
spike' but has many of the characteristics associated with impact cratering.
We also note that atoms on the rim borders are aligned along the
preferential [110] directions indicating recrystallization of the material
pushed or deposited onto the surface. For this ``atomic'' material a few
adatoms are also seen far from the track region. 
%%  These are mostly redeposited atoms which almost entered the gas phase, and were 
%%  readsorbed some distance from the track center. Few of them are atoms pushed 
%% out by the shock.}

\begin{figure}[tbp]
\centerline{\psfig{figure=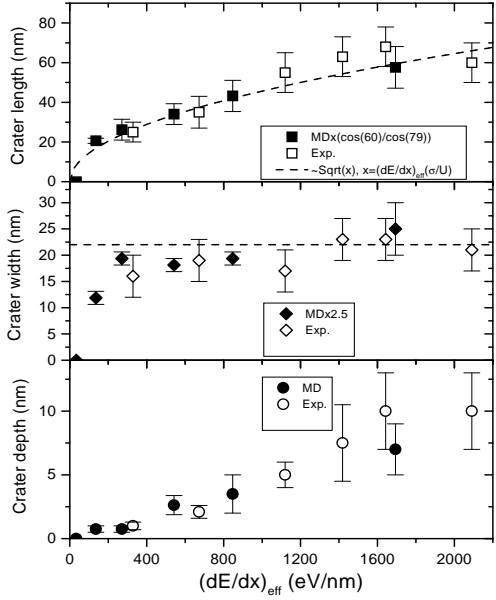,width=7.5cm,height=9.5cm,angle=0}}
\caption{Crater size as a function of $\left( dE/dx\right) _{eff}$. MD
results for $\Theta =60$, $U=0.5$ eV, $\sigma =5$ \AA , and taking $\left(
dE/dx\right) _{eff}=0.2\left( dE/dx\right) $. MD crater length is multiplied
by a factor $\left( \cos 60/\cos 79\right) $ to account for the different
incident angle in the simulation. Open symbols are experimental data from
Papal\'{e}o {\it {et al} {\protect\cite{papaleo}}. }}
\label{fig3}
\end{figure}

The crater formation has several stages, but most of the crater volume is
ejected before $\sim 20t_{o}$ (35 ps for a mass of $70u$). Temperature
varies greatly during the formation process and near the center of the track
it can be larger than the melting temperature even after $10t_{o}$. The
dependence of the crater dimensions on the energy deposited in the track can
be seen in Fig. \ref{fig3}. The MD values represent the mean value of the
crater dimension for 4-8 simulations at each $\left( dE/dx\right) _{eff}$,
and the maximum difference among the mean value and values for particular
simulation was taken as the error bar. Because of sample size limitations,
angles above 60 degrees are problematic whereas experimental results are
often performed at nearly grazing incidence.\ In previous work \cite{angle},
the width of the distribution of original position of the ejecta along the
direction of the incident beam was found to have a $\cos ^{-1}\Theta $
dependence while no variation was found in the distribution along the
perpendicular direction. Therefore, in order to compare the simulations with
the experiments a $\cos ^{-1}\Theta $ dependence has been assumed for
lengths along the direction of incidence of the ion. As mentioned
before, the MD results for crater and rim length obtained at $\Theta =60$
were multiplied by a factor $\cos 60/\cos 79$ when comparing to the data of
Papal\'{e}o {\it et al.} \cite{papaleo}. Experimental results are the
convolution of the actual crater profile with the AFM tip shape, and may
also involve some late relaxation of the crater walls. Therefore the
measured depth is expected to be smaller than the actual depth. The effect
of the convolution is smaller for the other crater dimensions. Finally, the
initial track size is not only difficult to estimate but may vary with ion
velocity. We have used $r_{cyl}=$ 10 \AA\ in our simulations, but it is not
unreasonable that the initial spike radius is 5 times larger. For all these
reasons the MD results were normalized separately for the length, width and
depth comparisons in Fig 3. However,\ it is seen, quite remarkably, that 
{\bf the trends in the experiment and in the simulation are the same}. This
indicates that useful scaling laws can be obtained and that the crater
formation process is insensitive to the details of both the energy
deposition profile and the materials properties. This also means that
simulations for relatively simple systems can be used to predict crater
structures when using ions to modify materials.

\begin{figure}[htb]
\centerline{\psfig{figure=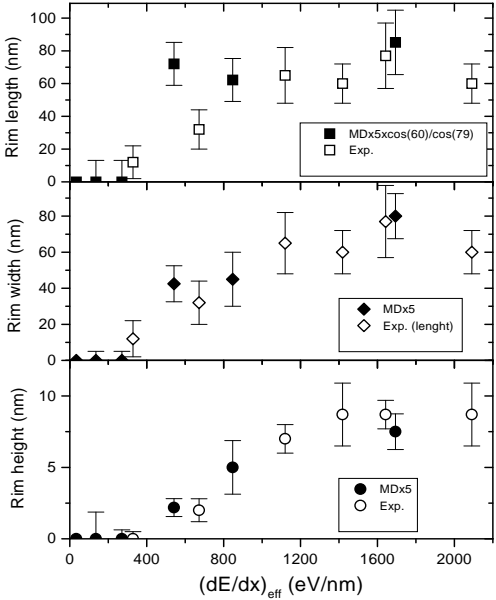,width=7.5cm,height=9.5cm,angle=0}}
\caption{Same as Fig. 3, but showing rim size as a function of $\left(
dE/dx\right) _{eff}$. MD sizes are multiplied by 5 in order to compare
trends with the experiment.}
\label{fig4}
\end{figure}

For the Yamamura et al. \cite{yamamura,yamamura1} shock model given in the appendix,
the volume of ejecta is determined by the energy density deposited, and the
crater length and width are {\bf both} expected to increase as $\sqrt{\left(
dE/dx\right) _{eff}}$. Assuming that the volume removed and the crater size
are directly related, this gives a $\left( dE/dx\right) _{eff}^{3/2}$
dependence for the sputtering yield. The crater length is seen in Fig. 3 to
increase roughly as $\sqrt{\left( dE/dx\right) _{eff}}$ in both the
simulations and the experiments. However, the length appears to increase
more slowly and to saturate at large values of $\left( dE/dx\right)
_{eff}$. After a steep increase, the crater width is seen to be much smaller
than the length and increases only very slowly with increasing $\left(
dE/dx\right) _{eff}$. Crater depth is much smaller than both the length and
the width but appears to increase with $\left( dE/dx\right) _{eff}$. The
pressure pulse model, also discussed in the appendix, gives such a scaling,
but it also predicts that all of the dimensions have the same scaling. That
model gave a good fit to MD calculations of the sputtering yield for
ejection of large L-J molecules with hard cores, at normal incidence, and
appeared to agree with data from a solid made of large biomolecules \cite
{david}. However, it differs from what is seen in the simulations
presented here and in the polymer experiments.

Although it is not clear from Fig. 3, for $E_{exc}$ below $U$ (i.e., the
energy density in the track is less than the cohesive energy density) no
crater is formed. Such a threshold is also seen in the experiments and,
therefore, crater detection can give a measure of the cohesive energy. Below
the threshold, several atoms escape from the top layers leaving vacancies 
and the sputtering yield is small, as discussed by Bringa {\it et al.} 
\cite{paperII}. In addition, some atoms are displaced to the top layer,
where they stay as adatoms, but an identifiable crater is not observed. For 
$6U>E_{exc}>2U$ a shallow crater forms and again several atoms are 
re-located as adatoms on the surface. For $E_{exc}>6U$ the energy density is 
about a tenth of the bulk modulus of the material, slip dislocations appear 
and a crater rim is formed. The dimensions of the rim are shown in 
Fig. \ref{fig4} as a function of $\left( dE/dx\right) _{eff}$. After the 
initial rise at `threshold' the rim length and width stay constant within 
our error bars, but the rim height increases very slowly with 
$\left( dE/dx\right) _{eff}$. Again, these trends are also observed 
experimentally, which is quite remarkable considering the differences 
in materials.

\section{Sputtering yield}

\begin{figure}[htb]
\centerline{\psfig{figure=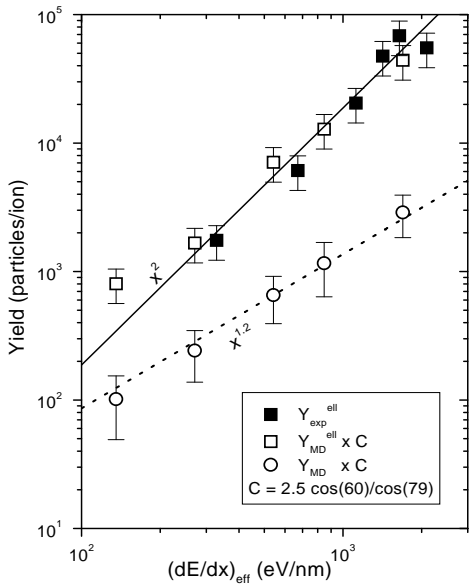,width=7.5cm,height=9.5cm,angle=0}}
\caption{Sputtering yield as a function of $\left( dE/dx\right) _{eff}$.
Yields obtained assuming the crater is an ellipsoid: $Y_{exp }^{ell}$ from
experiments of ion bombardment of polymers {\protect\cite{papaleo} }(solid
squares), $Y_{MD }^{ell}$ from the size of MD craters in Fig. 3 (open
squares). The ``true'' MD yields,$Y_{MD}$, are also shown (open circles).
Lines are only a guide to the eye.}
\label{fig5}
\end{figure}

The sputtering yield can, of course, be obtained directly from the MD
simulations. We showed earlier the surprising result that at the high
excitation densities for which the energized track produces craters, the
sputtering yield, $Y_{MD}$, is not predicted by standard models. 
$Y_{MD}$ is roughly proportional to $\left( dE/dx\right) _{eff}$ times the 
effective `sputter depth', which is a fraction of the initial track width 
\cite{paperII,angle}. When it is difficult to measure it directly, the 
sputtering yield is often approximated by an estimate of the crater volume. 
This has been tested for normal incidence for L-J molecules with a core 
\cite{david}. Here we evaluate that procedure for large incident angle for 
a standard L-J solid. Typically one assumes that the ejected volume is a
semi-ellipsoid, $Y^{ell}=\left( \pi /6\right) nr_{lc}r_{wc}z_{c}$, where 
$r_{cl}$, $r_{cw}$ and $z_{c}$ are the crater length, width, and depth
respectively. This rough estimate is based on crater shapes obtained from MD
and has been used in several papers \cite{papaleo}. In Fig. \ref{fig5} we
show $Y^{ell}$ for the polymers \cite{papaleo} ($Y_{exp }^{ell}$) and the MD
crater ($Y_{MD}^{ell}$) using the MD values for the crater in Fig. 3. These
are both compared to the actual MD yield, $Y_{MD}$, as a function of $\left(
dE/dx\right) _{eff}$.

The polymer yield was obtained as in Papal\'{e}o {\it et al.} \cite{papaleo}
using a semi-ellipsoidal volume and the same number density as in the MD
simulations (using $M=70u$ gives a density of $0.98$ g/cm$^{3}$). The
experimental yield estimate for polymers is larger than the `yield'
estimated from the MD craters. The MD yields have been multiplied by a
constant $C$, $C=2.5\times \cos \left( 60\right) /\cos \left( 79\right) $,
which results from the normalization of crater size in Fig. 3. However, the
trends are the same. They both give a yield that depends quadratically on $%
\left( dE/dx\right) _{eff}$. This is, fortuitously, the same dependence
predicted by thermal spike models for sputtering. However, it is seen that
these yield estimates are an order of magnitude larger than $Y_{MD}$, and
that they have a steeper dependence on $\left( dE/dx\right) _{eff}$. This
large discrepancy is due to several factors. First, the craters are not well
approximated by a half ellipsoid. Second, for the excitation energies shown
many of the atoms originally in the crater relocate on the rim. Third, there
are regions of higher density at the crater walls. These factors add up to a
surprisingly large overestimate of the experimental yield. More importantly,
since the ratio $Y_{MD}^{ell}/Y_{MD}$ changes with $\left( dE/dx\right)
_{eff}$, the dependence of the yield with $\left( dE/dx\right) _{eff}$ can
not be obtained from such estimates. Since most of the atoms on the rim come
from the crater region, a more detailed description of the morphology of the
crater needs to be made to obtain a reasonable yield estimate.

\section{Discussion and Summary}

Here we carried out a series of MD simulations to study crater formation due
to the high energy density deposited in a cylindrical `track'. Such an
energized track might be formed by a penetrating fast ion that deposits its
energy in electronic excitations, which is of interest here, or deposits its
energy by momentum transfer producing recoil atoms. That is, the craters
described are not impact craters like the lunar craters \cite{moon},
rather they are the craters formed in response to the rapid energy
deposition in the track of an energetic ion. The study here is for large
angles of the track with respect to the surface normal.

First, we showed that the crater structure remains stable in this
Lennard-Jones material over the longest simulations time tested ($\sim
75t_{o}$), which is much larger than the crater formation time ($\sim
20t_{o} $). This is the case because of the rapid cooling of the track by
the melting and pressure pulse processes described in earlier papers \cite
{paperII,prb-new}. We also found the initially surprising result that the
scaling of the crater parameters with $\left( dE/dx\right) _{eff}$ in this
L-J solid agrees remarkably well with that found experimentally, for MeV
heavy ion bombardment of polymers at 79 degrees to the normal \cite{papaleo}%
. This means that concepts learned from MD simulations of simple materials
can be applied to more complex materials.

As shown earlier for the experimental data for polymers \cite{papaleo}, we
find here a threshold for crater formation and a second threshold for rim
formation. In another set of experiments the rims could be removed when the
polymer is maintained at higher temperatures so that viscous relaxation
occurs. Therefore, Papal\'{e}o et al. \cite{ricardoprb} used the relaxation
of rim formation vs. material temperature to locate the glass transition
temperature. Here we did not vary the material temperature as late
relaxation occurring over long time periods can not be described using MD.
However, in this paper we are able to relate the two thresholds to the track
energy density. Our MD simulations show that the threshold for crater
formation occurs when the energy density in the track is close to the
cohesive energy density or, in the track formation model used here, when the
non-radiative relaxation energy per particle inside the initial track, $%
E_{exc}$, is near the sublimation energy $U$. The threshold for rim
formation, however, occurs at a higher energy density both in experiment and
in the simulations. We find this to be, $E_{exc}\approx 6U$, which occurs
when the energy density in the track roughly equals the bulk modulus of the
material. In the crystalline material slip dislocations can form at such
energy densities allowing the raised structure to be maintained. Using an
efficiency of 0.2 for determining $\left( dE/dx\right) _{eff}$ and our
simulation parameters, we find that the experimental value of the stopping
power needed to form a rim would be around $2$ keV/nm for $\Theta=79$.

Above the threshold the crater width is found to be nearly constant for
large incident angles and the crater length and depth increase sub-linearly
with $\left( dE/dx\right) _{eff}$. The rim height is ten times smaller than
its length and grows faster than the crater size (depth and length). This
dependence should be compared to the steeper dependence for the crater
radius and depth at normal incidence \cite{reimann-crater}.

We note that for non-penetrating cluster bombardment the crater scaling with
energy deposited is different from that found here. For incident clusters
the energy of the projectile is deposited close to the surface. For a
projectile energy $E$, the crater volume, $V$, is found to follow \cite
{insepov2000} $V\propto \left( E/U\right) $. However, recent MD simulations
seem to indicate that for keV copper clusters on copper scaling is \cite
{urba-crater} $V\propto (E/U^{2})$. \ It was argued that the presence of a
molten region in the MD simulation caused the steeper dependence on $U$. New
MD results from simulations of keV xenon ions on gold \cite{kai2001} support
the quadratic dependence with $U$ and relate this to the formation of a
melt. It is difficult to compare these results with ours in which the energy
is deposited in a long cylindrical track. Assuming that the energy relevant
for crater formation is deposited in a volume close to the surface of depth 
$L$, $E=\left( dE/dx\right) _{eff}L$, we find $V\propto \left( E/U\right)^{2}$. 
Here we also find that the molten region is important in the crater
formation. Clustering of the ejecta might also affect crater size \cite
{reimann-crater}, but in our simulations, unlike the EAM\ Cu used in several
cluster bombardment simulations \cite{urba-crater}, there is almost no
contribution of clusters to the sputtering yield.

Finally, we examined the accuracy of roughly estimating the sputtering yield
by simply parametrizing the crater volume. Recent results by Insepov {\it et
al.} \cite{insepov2000} point to a possible connection between crater size
and hardness, and claim that crater volume is also related to the sputtering
yield. However, they find a different dependence on the bombarding energy
for the yield ($Y\propto E^{1.4}$) and for the crater size ($V\propto E$),
confirming the discrepancy found here ($Y\propto \left( dE/dx\right)
_{eff}^{1.2}$ and $V\propto \left( dE/dx\right) _{eff}^{2}$). We showed
that, for the model material studied here, using crater size to estimate
sputtering yield can produce surprisingly large errors in the sputtering
yield and, even, the wrong dependence on $\left( dE/dx\right) _{eff}$.
Therefore, if the yield can not be measured directly, the full morphology to
the track and rim need to be described to get an accurate yield.

\acknowledgments
This work was supported by the National Science Foundation Astronomy and 
Chemistry Divisions. We would like to thank useful comments from H. M. 
Urbassek, K. H. Nordlund, and Z. Insepov. E. Hall, from the Research Computing 
Support Group at UVa., helped creating the contour maps in Fig. 2.

\appendix

\section{Shock Models for sputtering and crater formation}

There are several related models which attempt to explain sputtering at high
excitation density using shock waves. Also, the collision of an impactor
with a target producing spallation has been extensively studied with MD \cite
{holian1}, together with cluster bombardment-induced shock waves \cite
{insepov,moseler}. The spallation process originates from the interaction of
two rarefaction waves, one coming from the shock wave reflected at the
surface and the other coming from the impactor \cite{shock}. Yamamura and
coworkers \cite{yamamura,yamamura1} estimated the sputtering yield due to
shock waves with spherical symmetry intersecting a surface. They suggested
that a hemispherical volume is ejected with radius$\,r_{c}$. Then the yield, 
$Y$ is proportional to the volume of the ejecta $(\sim 2\pi r_{c}^{3}/3)$
with $r_{c}\sim \left( dE/dx\right) _{eff}^{1/2}$ . Bitensky and Parilis 
\cite{bitensky} considered cylindrical tracks and the incident angle
dependence to model biomolecule sputtering. At normal incidence their model
reduced to the spherical shock model. The crater dimensions are assumed to
be proportional to $\sqrt{\left( dE/dx\right) _{eff}}$.

In order to explain experiments on ejection of whole biomolecules \cite
{david} where $Y\propto \left( dE/dx\right) _{eff}^{3}$, Johnson {\it et al.}
\cite{david} proposed the pressure pulse model (PP). In the PP\ model, there
are many excitation events along the ion track, each contributing to $\left(
dE/dx\right) _{eff}$ as in the simulations described here. Whereas the
energy density evolves diffusively the net energy density gradient causes a
net volume force and, therefore, a net momentum transfer radially and
towards the surface. If the net momentum transfer to a certain volume is
larger than some critical momentum that volume will be ejected. This
determines a critical radius $r_{c}\propto (dE/dx)_{eff}$, with the volume
ejected proportional to $r_{c}^{3}$. The PP model predicts an angular
distribution peaked at 45$^{o}$ and agrees well with MD\ simulations that
use a Lennard-Jones potential with a core to describe the interactions of
large excited molecules \cite{david}. Notice that the PP\ model gives crater
dimensions proportional to $\left( dE/dx\right) _{eff}$. If a critical
energy for ejection is considered \cite{bitensky1}, instead of a critical 
momentum, the yield is $Y\propto \left( dE/dx\right) _{eff}^{3/2}$, as in
Kitazoe {\it et al. }\cite{yamamura}.

In all models discussed above the width and length of the crater have the 
{\bf same} dependence on $\left( dE/dx\right) _{eff}$. When oblique ion
incidence is considered the yield increases as $1/\cos \Theta $ because the
length of the crater increases also as \cite{reimann-crater} $1/\cos \Theta $%
. Therefore, the $\left( dE/dx\right) _{eff}$ dependence is the same as at
normal incidence, which is not what is found in the simulations presented
here.

\end{multicols}


\begin{references}
\bibitem{binding}  The bulk modulus, which is known from experiments, can be
taken proportional to the sublimation energy.
\end{references}
\end{document}